# Hybridized magnetic microwire metacomposites towards microwave cloaking and barcoding applications

Y. Luo[1,a)], F.X. Qin[2,a)], F. Scarpa[1], J. Carbonell[3], M. Ipatov[4], V. Zhukova[4], A. Zhukov[4], J. Gonzalez[4], H.X. Peng[2,b)]

[1]*Advanced Composites Centre for Innovation and Science, Department of Aerospace Engineering, University of Bristol, University Walk, Bristol, BS8 1TR, UK*

[2]*Institute for Composites Science Innovation (InCSI), School of Materials Science and Engineering, Zhejiang University, Hangzhou, 310027, PR China*

[3]*Wave Phenomena Group, Universitat Politècnica de Valencia, Camino de Vera, s/n, 46022 Valencia, Spain*

[4]*Dpto. de Fisica de Materiales, Fac. Quimicas, Universidad del Pais Vasco, San Sebastian, 20009, Spain*

The microwave behavior of polymer metacomposites containing parallel Fe-based and continuous/short-cut Co-based microwire arrays has been investigated. A magnetic field-tunable metacomposite feature has been identified in the dense continuous hybrid composite confirmed by the transmission windows in the frequency band of 1 to 3.5 GHz. The complex magnetically tuned redshift-blueshift evolution of the transmission window is reasoned to result from the competition between the dynamic wire-wire interaction and the ferromagnetic resonance of Fe-based wires. Increasing Co-based inter-wire spacing to 10 mm in the continuous hybrid composites, a remarkable dual-band transmission window in the 1.5-3.5 GHz and 9-17 GHz is respectively induced by the ferromagnetic resonance of Fe-based wires and the magnetic resonance arising between Fe-Co wire couples. The hybridization of parallel Fe-based and short-cut Co-based wires in the polymer composite leads to a significant enhancement of the transmission window in the frequency band of 1 to 6 GHz due to the band-stop nature of Co-based wires. The advanced hybridized microwire metacomposites are arguably demonstrated to be particularly attractive for microwave cloaking and radio frequency barcoding applications.

---

a) Luo and Qin as joint first authors with equal contribution to the work.
b) Corresponding author: hxpengwork@zju.edu.cn(HXP)

# I. INTRODUCTION

Metamaterials have stimulated tremendous fundamental and practical interests in recent years owing to their peculiar electromagnetic (EM) properties and engineering applicability in a wide range of operating frequencies. Based on their artificial double negative (DNG) properties, i.e., negative permittivity and negative permeability, myriads of concepts and potential applications in the metamaterial family have been derived including invisibility cloaking,[1] perfect metamaterial absorbers,[2] perfect lenses,[3] mechanical metamaterials[4,5] since the first experimental demonstration achieved by the split ring resonators (SRRs) and metallic wires.[6] To date, enormous efforts have been devoted to designing multi-band[7,8] and band-tunable metamaterials[9], and the integration of mechanical and EM metamaterials[10] in order to maximize or optimize their EM performance. However, the realization of conventional metamaterials utilizes the overall dielectric and magnetic response of the mesostructure but probes less consideration of intrinsic properties of building blocks. Moreover, it also remains an issue of the intensively increased metamaterial-system complexity arising from the necessity of delicately deploying related components to become arranged conventional meta-structures, not to mention the difficulty placed in the manufacturing stage and high costs incurred therein.

Most recently, we have realized a metacomposite with the DNG characteristics, which is comprised of the continuous Fe-based amorphous ferromagnetic microwire arrays and aerospace-grade polymer-based composites.[11] This design provides a much simpler metamaterial structure and, most essentially, promises that DNG features can be realized in a 'true' material from an engineering standpoint rather than a merely functionalized structure. As is well known, Co-based microwires have the distinguished giant magnetoimpedance (GMI) effect as opposed to Fe-based wires due to their near-vanishing magnetostriction,[12] and their EM performance can be accordingly modulated by external magnetic fields/stresses.[13,14] It is therefore plausible that by incorporating the Co-based microwire arrays into the Fe-based wire/polymer metacomposite-system, the external stimuli-controlled metacomposite behavior can be anticipated. Hence, the underlying physics of the proposed hybrid microwire composites interacting with incident microwaves are worth a thorough investigation. Moreover, we have introduced the 'critical' spacing in the composites containing Fe-based wire arrays and explicitly demonstrated that wire spacing could exert essential influences on the metacomposite features.[11] Therefore it is of much scientific interest to further explore how the Co-based wire spacing would influence on the microwave performance of the hybrid wire composites. In addition, additional magnetic resonance could be realized from the microwave interaction between the Co- and Fe-based wire arrays, taken into account their different wire alignment, intrinsic EM properties and geometrical sizes. This would be beneficial to expanding DNG operating frequencies or enhancing microwave absorption. In this work, we hybridize the short-cut and continuous Co-based wire array respectively into the Fe-based microwire metacomposite and investigate their microwave properties. Key findings are summarized as follows. (i) For the continuous wires-enabled composites with high concentration of Co-based wires, the left-handed microwave behavior is observed evidenced by the emerging transmission windows when the external field is larger than 300 $Oe$; the transmission window peak redshifts with field increasing up to 600 $Oe$ due to the strong long dipolar interaction between wires and subsequently blueshifts with fields further increasing to 3000 $Oe$, suggesting a field-tuned metacomposite behavior; (ii) a 'natural' dual-band metacomposite characteristic is revealed in absence of external bias when reducing the spacing of Co-based array to 10 mm in the continuous hybrid composites; (iii) a transmission-window/band-stop feature is identified in the hybridized composite system containing short-cut Co-based wires and continuous Fe-based wire arrays.

The remainder of the paper is organized as follows. After providing the experimental details in Section II, we will discuss in Section III the key results obtained from three categories of samples: composites with closely packed continuous Fe and Co-based wires (named as dense continuous hybrid composites for convenience), composites with relatively less loaded continuous Fe and Co-based wires (dilute continuous hybrid composites) and composites containing orthogonal continuous Fe-wires but hybridized with short-cut Co-based wires (short-cut hybrid composites). The relevant applications of interest are also proposed therein. The paper is finally concluded in Section IV.

## II. EXPERIMENTAL

In the present work, amorphous glass-coated ferromagnetic microwires $Fe_{77}Si_{10}B_{10}C_3$ (total diameter of 20 μm, glass coat thickness of 3.4 μm) and $Fe_4Co_{68.7}Ni_1B_{13}Si_{11}Mo_{2.3}$ (total diameter of 15 μm, glass coat thickness of 7 μm) were manufactured by the modified Taylor-Ulitovsky technique[15] and supplied by TAMAG, Spain. To investigate the influences of wire topological arrangement on the microwave behavior of the composites, the Co-based and Fe-based microwires were hybridized into the aerospace-grade 913 E-glass prepregs in two topological arrangements, i.e., parallel Co-based and parallel Fe-based wire array (Fig. 1(a)); short-cut Co-based and continuous orthogonal Fe-based wire array (Fig. 1(b)). To consider the effect of wire spacing on the composite EM properties, in the continuous hybrid composite the spacing of Co-based wire array is arranged as 10 mm and 3 mm, respectively. A standard polymer curing procedure was then followed after the wire embedding: the prepregs were heated at a rate of 2.4 $^o$C/min to 125 $^o$C and temperature was kept for 120 min before naturally cooling down to room temperature. The pressure was elevated to 0.62 MPa at a rate of 0.07 MPa/min and remained for 270 min before decreasing at 0.07 MPa/min per minute.[16] All the resultant composite samples have an in-plane size of 500×500 mm$^2$ and thickness of 1 mm. It should be noted that the Fe-based and Co-based microwire arrays were arranged in separate prepreg layers to avoid the large reflection loss induced by the physical microwire contact.[11] Besides, the in-plane spacing between the Co-based and Fe-based microwire arrays was intentionally mismatched by approximately 1 mm to minimize the undesirably high reflection caused by the wire superposition. In comparison, polymer composites containing continuous Co- or Fe-based microwire array, orthogonal Fe-based wire array and short-cut Co-based wire array with same wire spacing as in the hybrid wire composites in Fig. 1 were also fabricated by the same experimental protocol.

The microwave behavior of the microwire composites were examined by the free-space measurement in the frequency band of 0.9 to 17 GHz with the electrical component $E_k$ parallel to the glass fiber direction (Fig. 1).[17] To track the magnetic field tunable properties, an additional external dc magnetic bias up to 3000 *Oe* was applied along the glass fiber direction. The effective permittivity was extracted via the obtained *S*-parameters by an implanted computer program: Reflection/Transmission Epsilon Fast Model. Further details of this experimental setup can be found in Ref [18].

## III RESULTS AND DISCUSSIONS

### A. Dense continuous hybrid composites

The panel (a) to (i) of Figure 2 describe the transmission, reflection and absorption coefficients of the continuous 10 mm Fe-based, 3 mm Co-based containing composites and their hybrid composite, in the frequency band of 1 to 6 GHz. It is clear that the EM performance of the Fe-based microwire containing composite can be tuned with the presence of rather high external fields, i.e., larger than 1000 *Oe* (Figs. 2(b), (e), (h)). A transmission dip (Fig. 2(b)) associated with an absorption peak (Fig. 2(h)) is identified approximately at 2.5 GHz, which can be attributed to the ferromagnetic resonance (FMR),[11] as evidenced by the blueshift of resonance peaks according to Kittel's equation.[19] It is indicated that the absorption intensity reduces in the vicinity of FMR frequency with increasing magnetic field. For Fe-based microwires, there exists a trade-off between effective permeability and ferromagnetic resonance frequency, as derived from Snoek's law:

$$(\mu - 1)f_{\text{FMR}} = \tfrac{2}{3}\gamma 4\pi M_s, \tag{1}$$

where γ, $M_s$, $f_{FMR}$ and µ denote the gyromagnetic factor, saturation magnetization, the FMR frequency and permeability, respectively.[20] The observed blueshift frequency results in the reduction of effective permeability, hence of the microwave absorption. In contrast, the field-tunability of Co-based wire containing composite is easily saturated at rather low fields up to 600 *Oe* (Figs. 2(c), (f), (i)) due to the wires' excellent soft magnetic properties.[21] In particular, the absorption of composite containing Co-based wires increases more significantly with increasing external field compared to Fe-based wires (Fig. 2(i)). This is due to the long-range dipolar resonance between Co-based wires since the 3 mm wire spacing is below the critical spacing.[11,22] Interestingly, the noted interaction resonance peaks in the absorption spectrum redshift with increasing fields (Fig. 2(i)). This can be explained as follows. With increasing magnetic fields, the skin depth of the Co-based wires at the interaction resonance frequency is significantly decreased due to the increased µ according to Eq. 2 as written below, which is consistent with the observed enhanced absorption.

$$\delta_{\text{res}} = \sqrt{\frac{\rho}{\pi f_{int} \mu}}, \qquad (2)$$

where $\rho$ and $f_{int}$ are electrical resistivity and interaction resonance frequency, respectively.[23] This subsequently realizes in rather high eddy current loss. As such, the frequency of wire-wire interaction resonance is then reduced to compensate this loss. It should be noted that the 10 mm spacing in the Fe-based microwire composite is too large to induce such interaction excitation, which dictates the blueshift of FMR peaks.[11]

In another perspective, the composites containing single Fe-based wire array exhibits no DNG characteristic due to the poor plasma frequency (1.4 GHz)[11] (also confirmed in Fig. 3(a)) arising from the longitudinal domain structure of Fe-based wires.[24] Although the plasma frequency can be enhanced in Co-based wires containing composite spaced by 3 mm, the overall high reflection loss suppresses the DNG features (Fig. 2(f)). We expect a solution by introducing Co-based microwires into the Fe-based wire composite. Indeed, lower reflection loss is revealed due to the improved impedance match (Fig. 2(d)). Remarkably, a transmission window is displayed in the frequency band of 1 to 3.5 GHz for the hybridized wire array composite when the fields are larger than 300 *Oe*. This suggests that the field-tunable double negative features are constructed in the continuous hybridized Fe-/Co-based wire composite system, which are distinct from the previously reported natural DNG features controlled by a critical spacing in the single Fe-based wires containing composites.[11] With the external fields increasing, the transmission window peak experiences a redshift-blueshift evolution (Fig. 2(a)). From the above reasoning, it is implied that such effect is because that the long dipolar resonance dominates at low magnetic fields of 600 *Oe*, which is induced by the interaction between wire couples,[11,25] and the FMR of Fe-based prevails at higher fields than 600 *Oe*. Above all, this magnetic bias-tunable metacomposite behavior satisfies such working requirements of the microwave invisibility cloaking that can be activated or deactivated by conveniently exerting an additional magnetic field.

Figure 3 exhibits the effective permittivity of the pure Fe-based wire composite and the continuous hybridized composites without or with external magnetic fields up to 3000 *Oe*. In the presence of external fields, the permittivity variation of composite containing single Fe-based wire array spaced by 10 mm is rather limited (Fig. 3(a)) due to the formation of nanocrystalline phase on the microwires after curing process, which degrades the field-tunable properties to a large extent.[11,26,27] Besides, the $f_p$ of Fe-based wires is heavily reduced in contrast to the theoretical calculation.[28] The dielectric response of the continuous microwire composite system can be described as the dilute plasmonic behavior and interpreted via

$$f_p = \sqrt{\frac{c^2}{2\pi b^2 ln\left(\frac{b}{a}\right)}}, \tag{3}$$

where $c=3\times10^8$ m/s, *b* and *a* denote the vacuum light velocity, wire spacing and diameter, respectively. To provide an explanation for the observed compromised $f_p$,[11] we have introduced a term of effective diameter $a_{eff}$ to consider the circumferential domain volume of the microwires, which gives the actual contribution to their dielectric performance.[28] For the Fe-based wires, the $a_{eff}$ is much smaller than their actual diameter because their circular domain only occupies trivial portion of the whole domain volume. Hence the $f_p$ is heavily reduced (Fig. 3(a)) and Eq. 3 should be modified to

$$f_p^2 = \frac{c^2}{2\pi b^2 \ln\left(\frac{b}{a_{eff}}\right)}. \qquad (4)$$

For the hybridized microwire array composite, the permittivity magnitude bears a decrease-increase trend with the magnetic fields increasing (Fig. 3(b)). As the external fields are less than 1000 $Oe$, it can be seen that the absolute value of permittivity decreases because of the improved impedance match ($Z$), which subsequently compromises the effective $\varepsilon$, given that

$$Z = \sqrt{\frac{\mu}{\varepsilon}}. \qquad (5)$$

Further increasing magnetic bias to 3000 $Oe$, the permittivity dispersion becomes stronger due to the increased μ arising from the FMR of the Fe-based microwires, creating circular fields that greatly enhance dielectric response of microwires.[28] To this end, the $f_p$ of the hybridized wire composite can also be tuned by external fields due to this additional magnetically enhanced circular field and therefore can be significantly elevated up to approximately 4.5 GHz (Fig. 3(b)). Another influential factor impacting on the dielectric performance is the wire-wire interaction. It is believed that the inter-wire interactions with presence of external fields among the Co-Co and Co-Fe wire couples after the hybridization have provided essential offset to the limited $a_{eff}$ in the continuous hybrid composite system, thus enhancing the $f_p$ to some degree. This is realized from the fact that the circumferential magnetic response excited by the external fields of a single microwire is beneficial to expanding the $a_{eff}$ of its neighbor wires.[11]

It should be addressed that, due to the small wire amount in the present composite system, directly extracting effective permeability dispersion from the *S*-parameters is quite difficult due to the measuring limits of the free-space device. Thus only the effective permittivity is displayed herein. To provide more concrete evidence in terms of the double negative characteristic associated with the field-tunable behavior, we present in Fig. 4 the frequency plots of transmission phase of the hybrid composite with the presence of external magnetic bias. Clearly, it can be seen that the derivative of the phase function reverses its sign when magnetic fields exceed 300 *Oe*, implying the anomalous refractive index dispersion region due to the negative group delay.[29] Moreover, the identified negative refractive index region (2 to 4 GHz as displayed in Fig. 4) overlaps with the above discussed transmission window frequencies in addition to the negative permittivity dispersion (Fig. 3(b)). As such, it is sufficient to conclude that under the tuning effect of external fields, the negative permeability is also obtained. It is worth mentioning that negative group delay is also identified in our previous parallel[11] and orthogonal[25] microwire containing composites, which confirms that the metacomposite features can be derived from the transmission windows in the microwire-composite case.

### B. Dilute continuous hybrid composites

It has been mentioned above that this field-tuned metacomposite features are prohibited at high frequencies in the dense hybrid composites due to the rather high reflection loss arising from the narrow spacing between the Co-based wires. Hence, it is natural to realize that, by increasing Co-based wire spacing, a high frequency transmission window could be attained in the continuous hybrid composite system arising from the possible magnetic resonance between wire couples. It is exhibited the EM parameters of the composites containing 10 mm spaced Co- and Fe-based wire arrays and their hybridized arrays in the frequency band of 0.9 to 7 GHz in Fig. 5. First, the Co-based wire composite displays the weak transmission and field-tunable properties, which is consistent with the overall high reflection loss as shown in Fig. 5(b). This suggests that the 10 mm spacing in the single Co-based wire array would still be too narrow to maintain double-negative features in the measuring frequencies. However, strikingly, a transmission window is identified with the 10 mm spaced Co-based wire array added to the 10 mm spaced Fe-based microwire composite in the frequency band of 1.5 to 5.5 GHz without dc fields, experimentally proving the DNG characteristic of composite containing dilute hybrid wire arrays. It should be reminded that this behavior resembles that of Fe-based microwire composite with the natural DNG feature.[11] By introducing the additional Co-based wire array into the composite, the $f_p$ is enhanced due to the increase of the continuous wire medium complexity therein,[28] as also evidenced in Fig. 7, resulting in the negative permittivity dispersion in the lower frequency band. Meanwhile, the negative permeability should be maintained at the frequencies between NFMR frequency (2.5 GHz according to above analysis) and ferromagnetic anti-resonance frequencies (FMAR) (larger than the maximum of whole measuring frequencies) of the Fe-based wires. However, a slightly different window position where the metacomposite features emerge is noted, i.e., the transmission window initiates from 1.5 GHz rather than 2.5 GHz. This small deviation can be attributed to the non-uniformly distributed frozen-in stress during microwires fabrication[26,30] and the modification effect of the residual stress distribution from polymer matrix via interface during the curing process.[27] As is well acknowledged, Co-based microwires have a much smaller NFMR frequency compared to Fe-based ones but the resonance peak position can be effectively tuned by the external magnetic bias.[14,12] However, the invariant transmission window position with respect to external fields implies that the left-hand behavior is independent of the intrinsic properties of Co-based microwires at low frequency regime (Fig. 5(a)). This is due to the weakened dynamic coupling between the EM waves and the single Co-based microwire array arising from the large dielectric loss as seen in Fig. 5(b). One might expect to extend the transmission window to even higher frequencies by decreasing the spacing of adjacent Fe-based wires, in the hope of increasing $f_p$ and receiving a stronger magnetic excitation from the composite with higher wire content. However, further investigations on the hybridization of Fe-based wire arrays with spacing of 7 or 3 mm plus 10 mm Co-based wire array reveal that the transmission window is cancelled out (*not shown here*). This is because that both heavily loaded Co- and Fe-based wires show such significant reflectance that prohibits the transmission window.

Figure 6 displays the transmission, reflection and absorption coefficients of the continuous microwire composites in the frequency band of 7 to 17 GHz. Overall, the strong transmission and reflection of Co-based wire array composite persists, owing to the drastically dwarfed skin depth at higher frequencies and consequently large magnetic and dielectric losses.[21] A feature of note is that the transmission enhancement in addition to the reflection and absorption dips have been acquired in the frequencies of 9 to 17 GHz for the Fe-/wide spaced Co-based hybrid wire array composite, indicating a natural DNG feature at such high frequencies. An absorption peak at 8.5 GHz is also noted, which implies that the magnetic resonance occurred and is responsible for the identified high frequency EM wave-induced transparency. It was elucidated in our earlier study that decreasing wire spacing to 3 mm would induce strong dynamic wire-wire interactions with microwave and realize the long range dipolar resonance.[11] Hence, further reducing the wire spacing to 1 mm, i.e., the mismatched spacing between Fe-based and Co-based microwire arrays in the present study, would also generate such effect and arouse the noticeable interaction absorption. Moreover, the circumferential fields created by the coupling between the $E_k$ and the longitudinal anisotropic field of Fe-based microwire array in favor of FMR[31] can also overlap and interact with the circumferential anisotropic field of Co-based microwire array, thus enhancing the magnetic excitation. At this point, this interactive magnetic resonance indicates a negative permeability dispersion above the resonance frequency. The observed high-frequency transparency also suggests a negative permittivity linking to the enhanced $f_p$ that will be discussed later. Overall, this transmission increase achieved by proper wire misalignment/offset and the dynamic excitation from propagating EM waves broadens the metacomposite operating frequencies and provides essential guides for the metamaterial design. In another perspective, according to Kittel's relations,[19] the above two noted magnetic resonance peaks and transmission peaks should have blueshifted with increasing external field. However, the experimental results support that they seem to be independent of dc fields. This can be explained by the above mentioned degraded magnetic performance of the microwires during the curing cycle.[27,32] This issue could be eased through the proper pretreatment to microwires such as magnetic field[33] or current annealing[34] to anticipate better static and dynamic EM properties.

To understand the dielectric behavior of the identified dual-band metacomposite characteristics, Fig. 7 shows the effective permittivity of dilute composites containing Fe-based, Co-based and their hybrid wire arrays in the presence of external field up to 3000 $Oe$. Substituting $b$=10mm, $a_{Co}$=15.97 μm and $a_{Fe}$=16.6 μm (inner core diameter) into Eq. 3, we have the $f_{p,Co}$=4.9 GHz and $f_{p,Fe}$=4.8 GHz, respectively. Obviously, compared with the measured value for the single Fe- or Co-based microwire array composite (Fig. 7), the calculated $f_p$ is overestimated by the equation. By employing the modified Eq. 4,[11] the $f_p$ of Co-based wire array composite is believed to be slightly higher than that of Fe-based ones (but still much lower than the theoretical value), as confirmed in Fig. 7, because they have larger circumferential domain volume, hence larger $a_{eff}$. This overestimated $f_p$ fundamentally clarifies the issue why the left-handed features are not accessible in the single Fe- or Co-based microwire array composites. Clearly, a prominent increase of the $f_p$ is obtained with the hybridization of Fe- and Co-based microwire arrays in the composite, i.e., a negative permittivity dispersion in nearly whole measuring frequencies (Fig. 7(a)). As per Eq. 4, the effective diameter is closely associated with the intrinsic domain structure of microwires. In the present case, taking into account the small spacing (~ one prepreg thick, i.e., 0.25 mm) between the Co- and Fe-based wire layers, the two neighboring Fe- and Co-based wires can be regarded as a wire pair/couple that interacts with the microwave (Fig. 1), that is to say, the hybridized composite effectively consists of 50 wire couples with approximately 10 mm spacing. As such, the long range dipolar resonance arising from dynamic wire interactions can greatly enhance the dielectric excitation of the wire pair unit at higher frequencies, thus improving the $a_{eff}$ and consequently the $f_p$. In polymer metacomposites, engineering $f_p$ towards higher frequencies is always a critical task in the frontier of metamaterial designing. To reduce the wire spacing or to elevate the metamaterial building block's geometrical dimensions appears to be feasible solutions according to Eq. 3, but neither of these would effectively suppress the excessive reflection loss or the complexity in the fabrication process due to the massive amount of necessary functional units. This negative aspect even makes the established dilute medium model inapplicable.[35,28] The hybridization of 'dilute' microwire arrays in the microwire-composite system, as shown in the present work, paves a new path to enhance the DNG frequency band, which offers more degrees of freedom in the metacomposite design and manufacturing. In a brief summary, the Co-based wire spacing has profound effects on the microwave behavior of hybrid composites: (i) the magnetic field-tunable properties could only be preserved by significantly decreasing the spacing of Co-based wire array to 3 mm; (ii) increasing Co-based wire spacing can mitigate high reflection loss at high frequencies and induce additional magnetic interaction between Fe-Co wire couples, thus developing a high frequency transmission enhancement.

### C. Short-cut hybrid composites

At this point, we have discussed the influence of continuous Co-based wire arrays on the microwave response in the microwire-composite system. What will be turned out if we introduce the short-cut Co-based wire arrays? Figure 8 shows the EM parameters of polymer composites containing short-cut Co-based wire array, continuous orthogonal Fe-based wire array and their hybridized wire arrays respectively. As was elucidated in our latest research, the orthogonal microwire array can generate DNG features in the microwire-polymer system due to the significantly improved $f_p$ arisen from the additional dielectric contribution of the vertical wires.[25] However, it remains an issue to tune the identified transmission windows in such orthogonal structure since the intrinsic properties of the microwires have been modified during the curing procedure and are less intuitively sensitive compared to their as-cast states. In this study, we employ the short-cut Co-based wire arrays. With reference to the Fig. 8(b), the transmission spectrum of the composite containing single short-cut Co-based wires reveals a typical band-stop dispersion indicated by a sharp transmission dip and a reflection peak at approximately 6 GHz. Note that short wires behave as dipoles when collectively interacting with incident waves, the dipole resonance frequency is expressed as Eq. 6:

$$f_{dr} = \frac{c}{2l\sqrt{\varepsilon_m}}, \tag{6}$$

where $\varepsilon_m$ and $l$ denote the permittivity of matrix materials and wire length respectively.[36,18] Substituting $\varepsilon_m$=3 and $l$=15 mm in the present study, we obtain $f_{dr}$=5.8 GHz, which coincides with the identified resonance peak in Fig. 8(c). In other sense, the artificial microwave opaqueness in the short Co-based wires containing composite is induced by the wire configuration therein. Strikingly, this band-stop characteristic is maintained in the hybridized composite and enhances the identified transmission window in the 1 to 6 GHz as received in the orthogonal Fe-based wire array composite. Hence the short-cut Co-based wire array plays a synergistic role in enhancing the double negative features of the composite, as also verified by the similar absorption coefficients profile of composites containing orthogonal Fe-based wire array and hybridized wire array (Fig. 8(c)); this is due to the averagely low absorption of the Co-based wires in the metacomposite operating frequencies. Together, this structure-induced opaqueness associated with the metacomposite features realized from the introduced wire arrangements in the composite structure and intrinsic magnetic/dielectric properties of comprised building blocks provide a feasible idea to achieve engineering metacomposites with meaningful metamaterial characteristics as against to the purely structure-associated conventional metamaterials. From the perspective of the absorption spectra (Fig. 8(c)), the absorption peaks are identified at 2 and 6.5 GHz which are originated from the FMR of Fe-based wires and dynamic wire-wire interaction in the orthogonal Fe-based structure, respectively.[11] Apart from this, an additional peak at 8.5 GHz is noted in the hybridized wire composite due to the dynamic interaction between short Co-based wires and continuous Fe-based wires, indicating that the Co-based wires can strengthen the absorption of the composite.

The effective permittivity of the composites containing short-cut Co-based wires, orthogonal Fe-based wires and their hybrid wire arrays is displayed in Fig. 9. One can observe that the hybridized composite has stronger permittivity dispersion than the pure Fe- and Co-based wire composites due to the large dielectric response induced by the inserted short Co-based wire array.

Meanwhile, the composite containing short-cut Co-based wire array reveals a non-plasmonic behavior verified by all-positive values of $\varepsilon'$ in the whole measuring band due to the discontinuous alignment of microwire arrays. The $\varepsilon''$ peak at 6 GHz also accords with the transmission dip due to the dipole resonance that results in the metacomposite/band-stop features as displayed in Fig. 8.

It is worth remarking that, compared with the metacomposite containing Fe-based wires only, the presented hybridized metacomposites shows metamaterial characteristics that are more tunable by external magnetic field. Also, the involvement of Co-wire offers more degrees of freedom to the composite structure design and associated properties control. As discussed above, several effects, i.e., FMR in Fe-based wire, Fe-Fe wire, Co-Co wire and Fe-Co wire interactions are involved and the predominating mechanism varies at different frequency. A combination of coarse and fine control of metacomposite behavior is therefore readily available via manipulation of any spacing and arrangement that are involved in this composite system. In view of these merits, the present composite may indicate a significant application of radio frequency identification (RFID), in which structure polymer composites are heavily used. The RFID is a contactless data capturing technique using RF waves to automatically identify objects. The ever-increasing applications of RFID for commercial inventory control in warehouses, supermarkets, hospitals as well as in military friend-and-foe identification has resulted in considerable research interest on low-costs, long range sensors design. Conventional RF tags are usually achieved either by the printed spacing-filling curves[37] or capacitively tuned dipoles.[38] However, these tags consist of complicated-shaped structure and thus requiring large manufacturing costs. Furthermore, the undesirable parasitic coupling effect of these structures when interacting with EM waves makes the precise analysis of their EM performance rather difficult. Our present study herein proposes a kind of versatile composite containing microwire arrays of a simple structure. By incorporating these wire composites into the objects to be detected, each object will have a unique ID coded in these composites. Moreover, the recent unfortunate 'MH370' event calls for application of such multifunctional composites to identify civil airplanes by their distinguished microwave response, considering their DNG features in the radar frequencies. Such application appears to be more practical than cloaking invisibility at this stage, and importantly, of more civil significance for e.g., distinguishing civil from military planes or even identify every airborne vehicle.

## IV. CONCLUSIONS

In conclusion, we have performed a systematic study on the microwave properties of hybrid polymer composites containing Fe-based and/or Co-based in different topological manner. The particular influences of the continuous/short-cut Co-based microwire arrays on the microwave response of hybrid composites have been recognized in the following three aspects: (i) in the continuous hybrid composite containing 3 mm spaced Co-based wire array, the embedded continuous Co-based wire array can induce magnetic field-tunable metacomposite characteristics when the fields are higher than 300 *Oe*; the transmission windows display a redshift-blueshift trend with the increasing of external magnetic fields in the continuous hybridized composite due to the competition between the interaction between Fe- and Co-based wire arrays and the FMR of Fe-based wires; (ii) in the composite containing 10 mm spaced Co-based wire array, a dual-band DNG feature is revealed in 1.5-5.5 GHz and 9-17 GHz respectively. The additional transmission window displayed at high frequencies originates from the dynamic magnetic interaction resonance between Fe-Co wire couples. (iii) The composite containing short Co-based wires and orthogonal Fe-based wires manifests a metacomposite/stop-band feature. This arises from the microwave opaqueness of the used Co-based wire arrays in the mid-frequencies of the measuring range, which exerts a synergistic effect on the identified transmission windows. All these findings showcase unique advantages created by the hybridization of Co-based and Fe-based microwires, which make the proposed metacomposites a strong candidate for microwave cloaking and radio frequency barcoding applications.

**Acknowledgements**

Yang Luo would like to acknowledge the financial support from University of Bristol Postgraduate Scholarship and China Scholarship Council. FXQ is supported under the 'Hundred Talents Youth Program' of Zhejiang University.

**References:**

[1] J. B. Pendry, D. Schurig, and D. R. Smith, Science **312**, 1780 (2006).

[2] N. I. Landy, S. Sajuyigbe, J. J. Mock, D. R. Smith, and W. J. Padilla, Phys. Rev. Lett. **100**, 207402 (2008).

[3] J. B. Pendry, Phys. Rev. Lett. **85**, 3966 (2000).

[4] Z. G. Nicolaou and A. E. Motter, Nat. Mater. **11**, 608 (2012).

[5] J. H. Lee, J. P. Singer, and E. L. Thomas, Adv. Mater. **24**, 4782 (2012).

[6] R. A. Shelby, D. R. Smith, and S. Schultz, Science **292**, 77 (2001).

[7] I. J. McCrindle, J. Grant, T. D. Drysdale, and D. R. Cumming, Adv. Opt. Mater. **2**, 149(2013).

[8] H. Li, L. H. Yuan, B.Zhou, X. P. Shen, Q. Cheng, and T. J. Cui, J. Appl. Phys. **110**, 014909 (2011).

[9] C. L. Chang, W. C. Wang, H. R. Lin, F. J. Hsieh, Y. B. Pun, and C. H. Chan, Appl. Phys. Lett. **102**, 151903 (2013).

[10] D. Shin, Y. Urzhumov, D. Lim, K. Kim, and D. R. Smith, Nature **4**, 4084 (2014).

[11] Y. Luo, H. X. Peng, F. X. Qin, M. Ipatov, V. Zhukova, A. Zhukov, and J. Gonzalez, Appl. Phys. Lett. **103**, 251902 (2013).

[12] M. H. Phan and H. X. Peng, Prog. Mater. Sci. 53, 323 (2008).

[13] J. Carbonell, H. García-Miquel, and J. Sánchez-Dehesa, Phys. Rev. B **81**, 024401 (2010).

[14] H. García-Miquel, J. Carbonell, and J. Sánchez-Dehesa, Appl. Phys. Lett. **97**, 094102 (2010).

[15] V. S. Larin, A. V. Torcunov, A. Zhukov, J. Gonzalez, M. Vazquez, and L. Panina, J. Magn. Magn. Mater. **249**, 39 (2002).

[16]H. X. Peng, F. X. Qin, M. H. Phan, J. Tang, L. V. Panina, M. Ipatov, V. Zhukova, A. Zhukov, and J. Gonzalez, J. Non-Cryst. Solids **355**, 1380 (2009).

[17]D. Makhnovskiy, A. Zhukov, V. Zhukova, and J. Gonzalez, Adv. Sci. Technol. (Faenza, Italy) **54**, 201 (2008).

[18]D. P. Makhnovskiy, L. V. Panina, C. Garcia, A. P. Zhukov, and J. Gonzalez, Phys. Rev. B **74**, 064205 (2006).

[19]C. Kittel, Phys. Rev. **73**, 155 (1948).

[20]J. L. Snoek, Physica **14**, 207 (1948).

[21]F. X. Qin, H. X. Peng, and M. H. Phan, Mater. Sci. Eng. B **167**, 129 (2010).

[22]Y. J. Di, J. J. Jiang, G. Du, B. Tian, S. W. Bie, and H. H. He, Trans. Nonferrous Met. Soc. China **17**, 1352 (2007).

[23]F. X. Qin, H. X. Peng, N. Pankratov, M. H. Phan, L. V. Panina, M. Ipatov, V. Zhukova, A. Zhukov, and J. Gonzalez, J. Appl. Phys. **108**, 044510 (2010). also J. Appl. Phys. **108**, 07A310, 2011.

[24]M. Vázquez and A. Zhukov, J. Magn. Magn. Mater. **160**, 223 (1996).

[25]Y. Luo, H. X. Peng, F. X. Qin, M. Ipatov, V. Zhukova, A. Zhukov, and J. Gonzalez, J. Appl. Phys. **115**, 173909 (2014).

[26]F. X. Qin, Y. Quéré, C. Brosseau, H. Wang, J. S. Liu, J. F. Sun, and H. X. Peng, Appl. Phys. Lett. **102**, 122903 (2013).

[27]H. Wang, F. X. Qin, D. Xing, F. Cao, X. D. Wang, H. X. Peng, and J. Sun, Acta Mater. **60**, 5425 (2012).

[28]J. B. Pendry, A. J. Holden, W. J. Stewart, and I. Youngs, Phys. Rev. lett. **76**, 4773 (1996).

[29]O. F. Siddiqui, M. Mojahedi, and G. V. Eleftheriades, IEEE Trans. Antennas Propag. **51**, 2619 (2003).

[30]H. Wang, D. W. Xing, X. D. Wang, and J. F. Sun, Metal. Mater. Trans. A **42**, 1103 (2011).

[31]L. Kraus, Z. Frait, G. Ababei, and H. Chiriac, J. Appl. Phys. 113, 183907 (2013).

[32]H. Wang, D. W. Xing, H. X. Peng, F. X. Qin, F. Y. Cao, G. Q. Wang, and J. F. Sun, Scr. Mater. **66**, 1041 (2012).

[33]J. S. Liu, D. Y. Zhang, F. Y, Cao, D. X. Xing, D. M. Chen, X. Xue, and J. F. Sun, Phys. Status Solidi A **209**, 984 (2012).

[34]J. S. Liu, F. X. Qin, D. M. Chen, H. X. Shen, H. Wang, D. W. Xing, M. H. Phan, and J. F. Sun, J. Appl. Phys. **115**, 17A326 (2014).

[35]F. X. Qin and H. X. Peng, Prog. Mater. Sci. **58**, 183 (2013).

[36]F. X. Qin, Y. Quéré, C. Brosseau, H. Wang, J. S. Liu, J. F. Sun, and H. X. Peng, Appl. Phys. Lett. **102**, 122903 (2013).

[37]J. McVay, A. Hoorfar, and N. Engheta, IEEE Radio Wireless Symp. San Diego, CA, Jan. 17–19, 2006, pp.199–202.

[38]I. Jalaly and I. D. Robertson, Eur. Microw. Conf. Paris, France, Oct. 4–6, 2005, vol. 2, pp. 4–7.

**Figure captions:**

FIG. 1 (Color online) Schematic illustration of the hybridization of (a) continuous parallel Fe-based microwire array plus continuous Co-based microwire array and (b) orthogonal Fe-based microwire array plus short-cut Co-based microwire array.

FIG. 2 (Color online) Frequency plots of the transmission coefficients, $S_{21}$, of composite samples containing (a) hybrid wire arrays with 3 mm spaced Co-based wires, (b) pure Fe-based wires and (c) pure Co-based wires; the reflection coefficients, $S_{11}$, of composite samples containing (d) hybrid wire arrays with 3 mm spaced Co-based wires, (e) pure Fe-based wires and (f) pure Co-based wires; the absorption coefficients of composites containing (g) hybrid wire arrays with 3 mm spaced Co-based wires, (h) pure Fe-based wires and (i) pure Co-based wires.

FIG. 3 (Color online) Frequency plots of the real part of effective permittivity, $\varepsilon'$, of the composites containing (a) continuous Fe-based wire array spaced by 10 mm and (b) hybrid arrays with 3 mm spaced Co-based wires in the presence of magnetic fields up to 3000 *Oe*.

FIG. 4 (Color online) Frequency dependence of transmission phase variation of the dense continuous hybrid microwire composite in presence of different external magnetic fields

FIG. 5 (Color online) (a) Transmission, (b) reflection and (c) absorption coefficients of composites containing 10 mm spaced Fe-based wire array, the 10 mm spaced Co-based wire array and their hybridized wire arrays in the frequency band of 0.9 to 7 GHz.

FIG. 6 (Color online) (a) Transmission, (b) reflection and (c) absorption coefficients of the same composites as shown in Fig. 4 measured in the frequency band of 7 to 17 GHz.

FIG. 7 (Color online) Frequency plots of (a) real part $\varepsilon'$ and (b) imaginary part $\varepsilon''$ of effective permittivity of composites containing 10 mm spaced Fe-based wire array, the 10 mm spaced Co-based wire array and their hybrid wire arrays, respectively.

FIG. 8 (Color online) (a) Transmission, $S_{21}$ (b) reflection, $S_{11}$ and (c) absorption coefficients of composites containing orthogonal Fe-based wire array, short-cut Co-based wire array and their hybridized wire arrays in the 0.9 to 17 GHz frequency range.

FIG. 9 (Color online) Frequency plots of the (a) real part, $\varepsilon'$, and (b) imaginary part, $\varepsilon''$, of effective permittivity dispersion of composites containing orthogonal Fe-based wire array, short-cut Co-based wire array and their hybridized wire arrays.

**Figures (next page):**

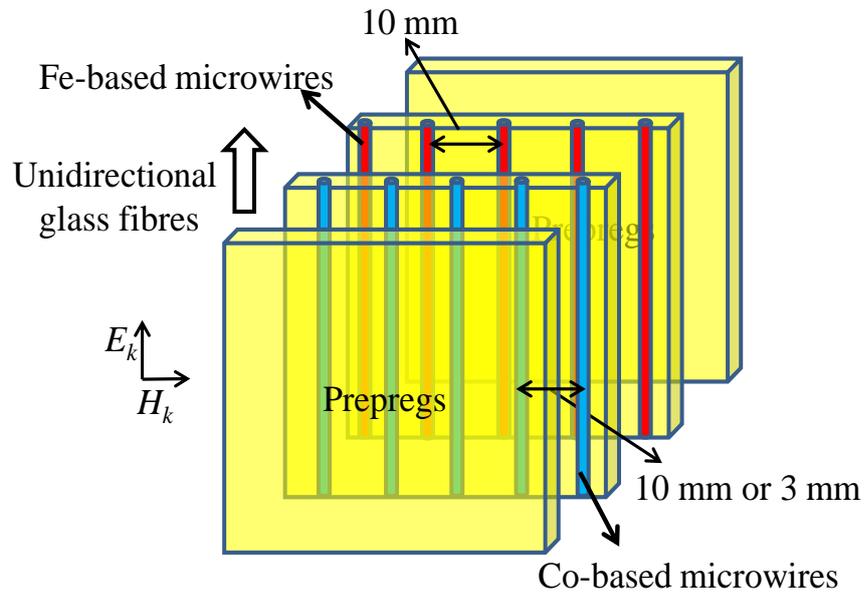
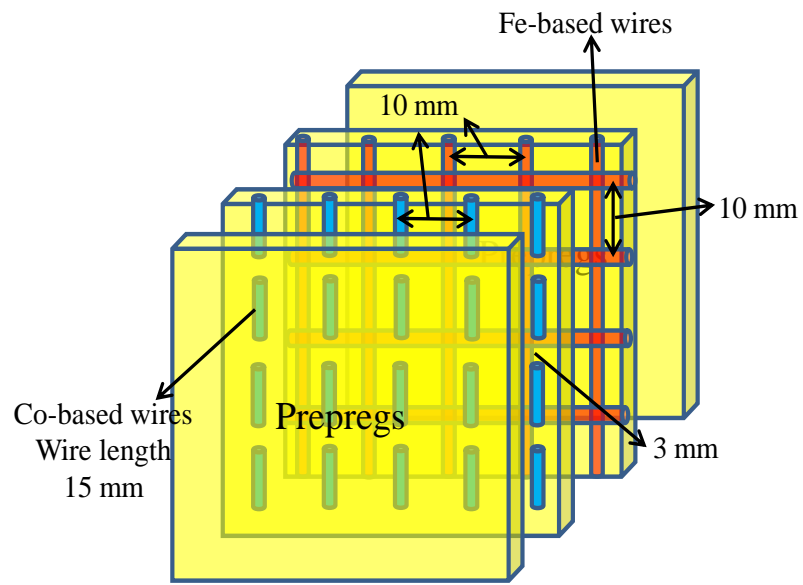

FIG 1

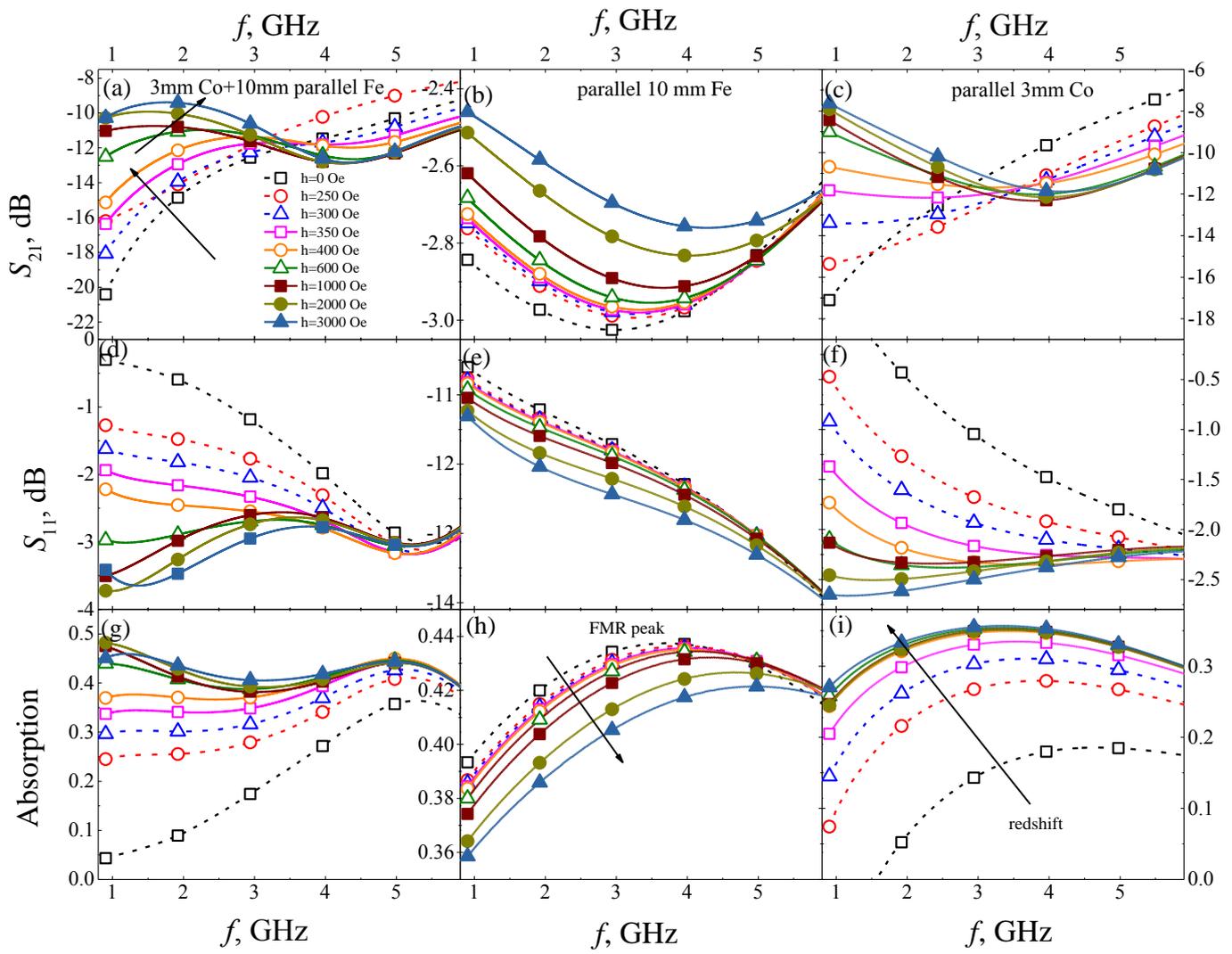

FIG 2

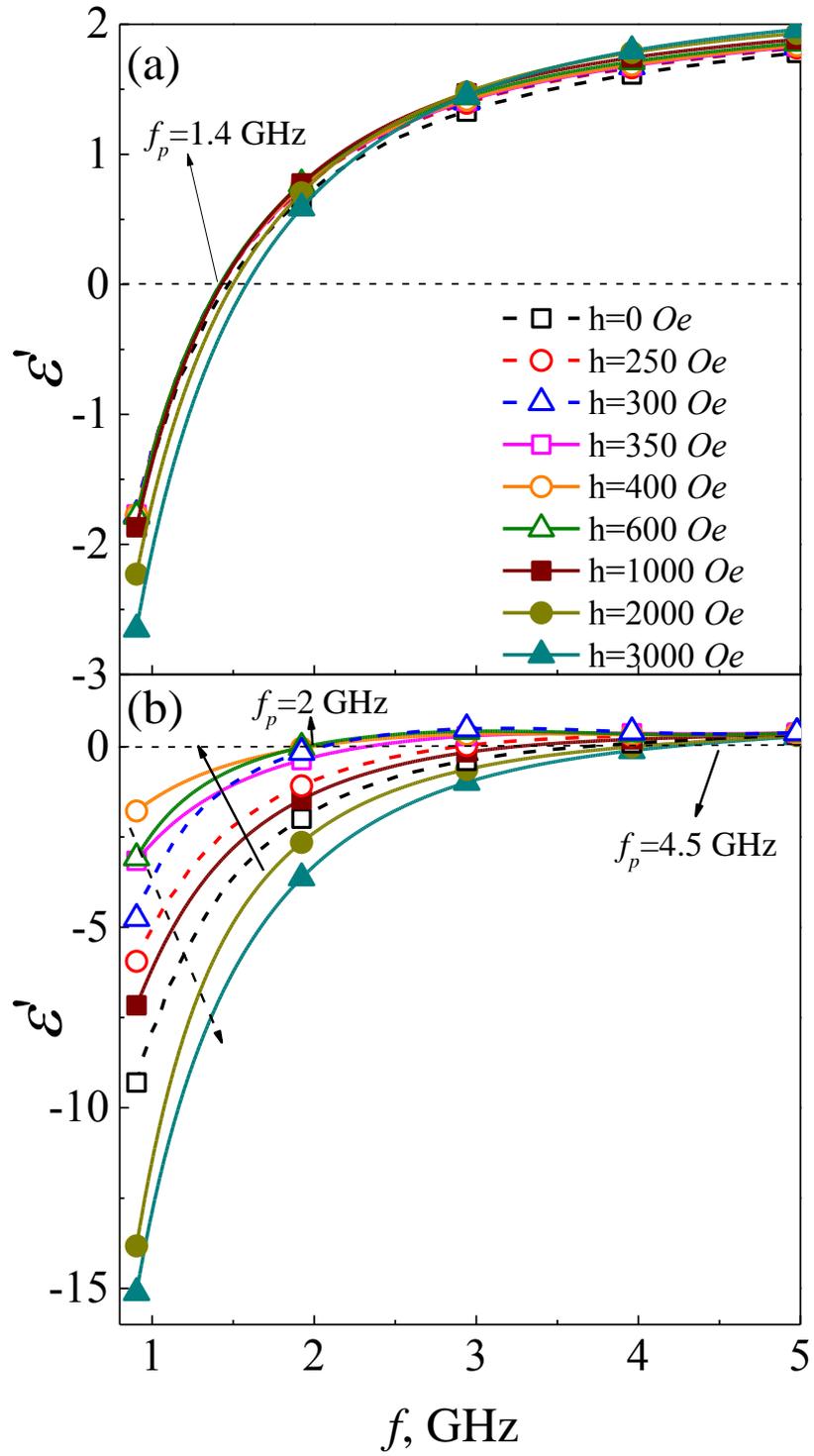

FIG 3

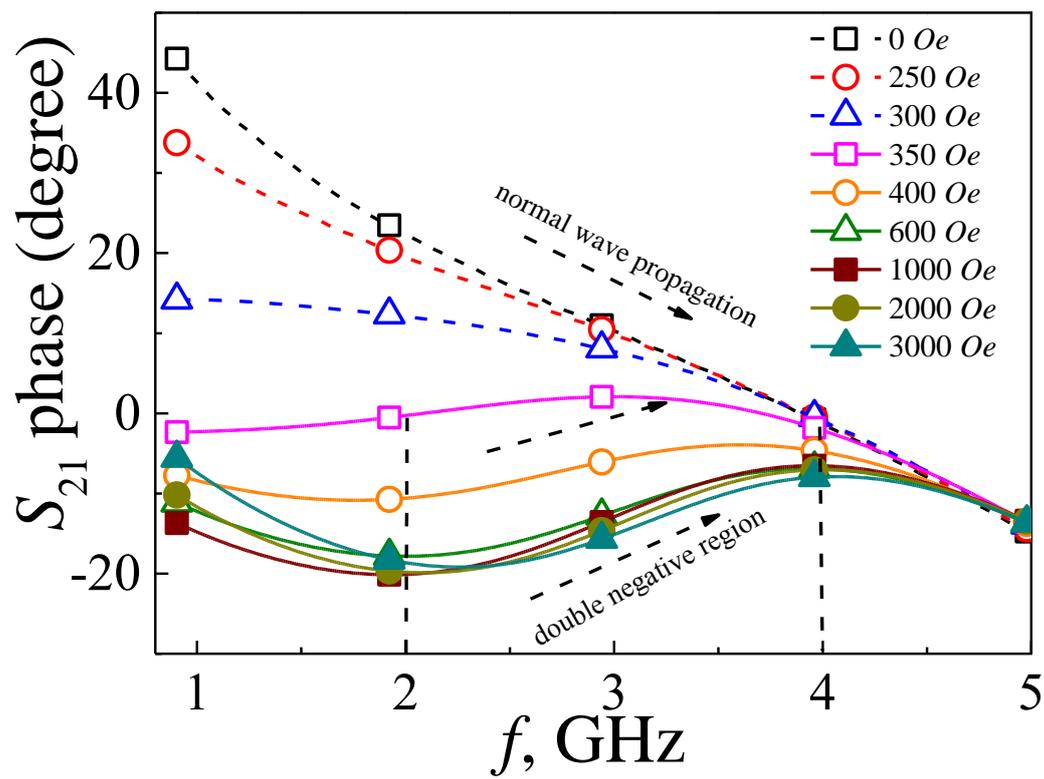

FIG 4

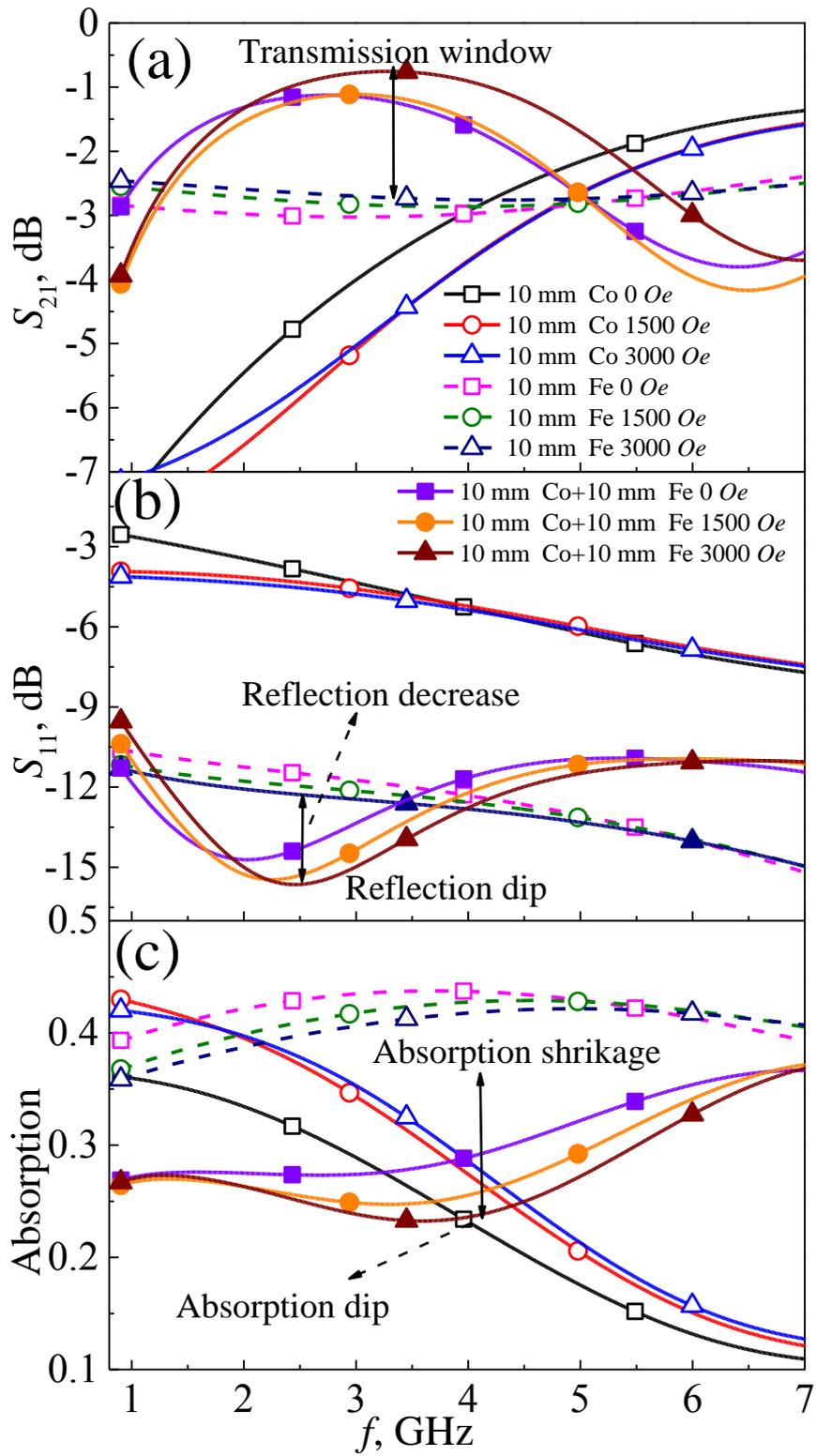

FIG 5

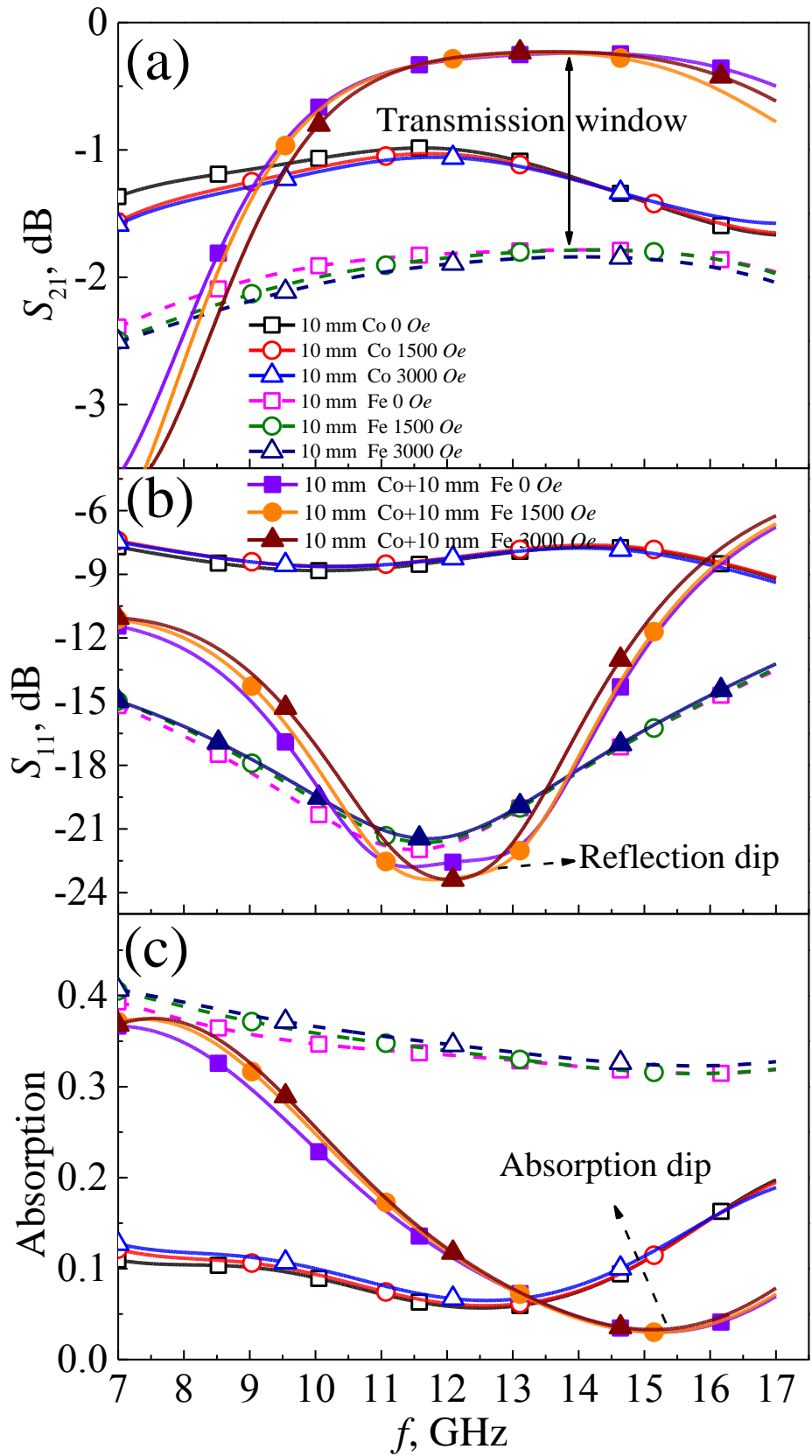

FIG 6

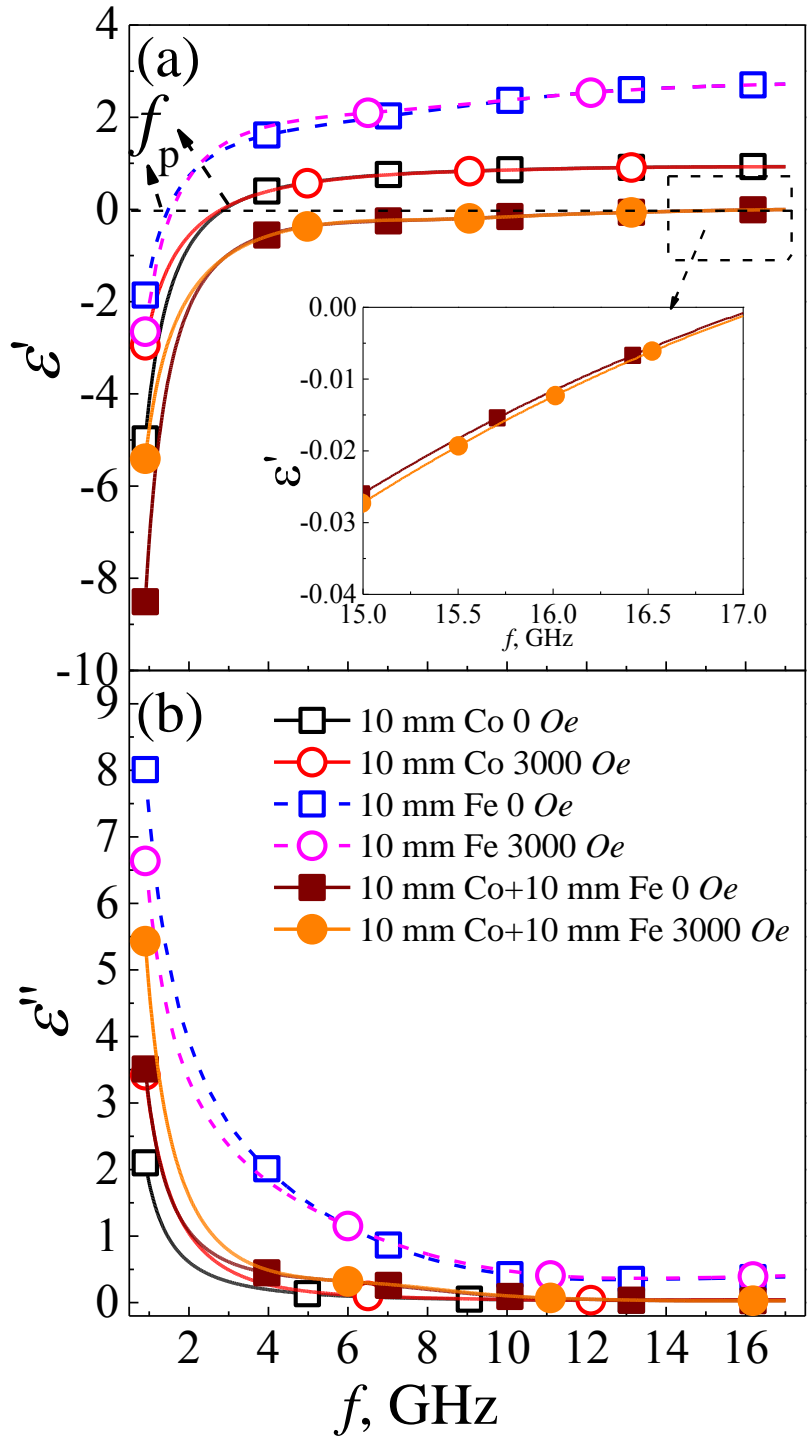

FIG 7

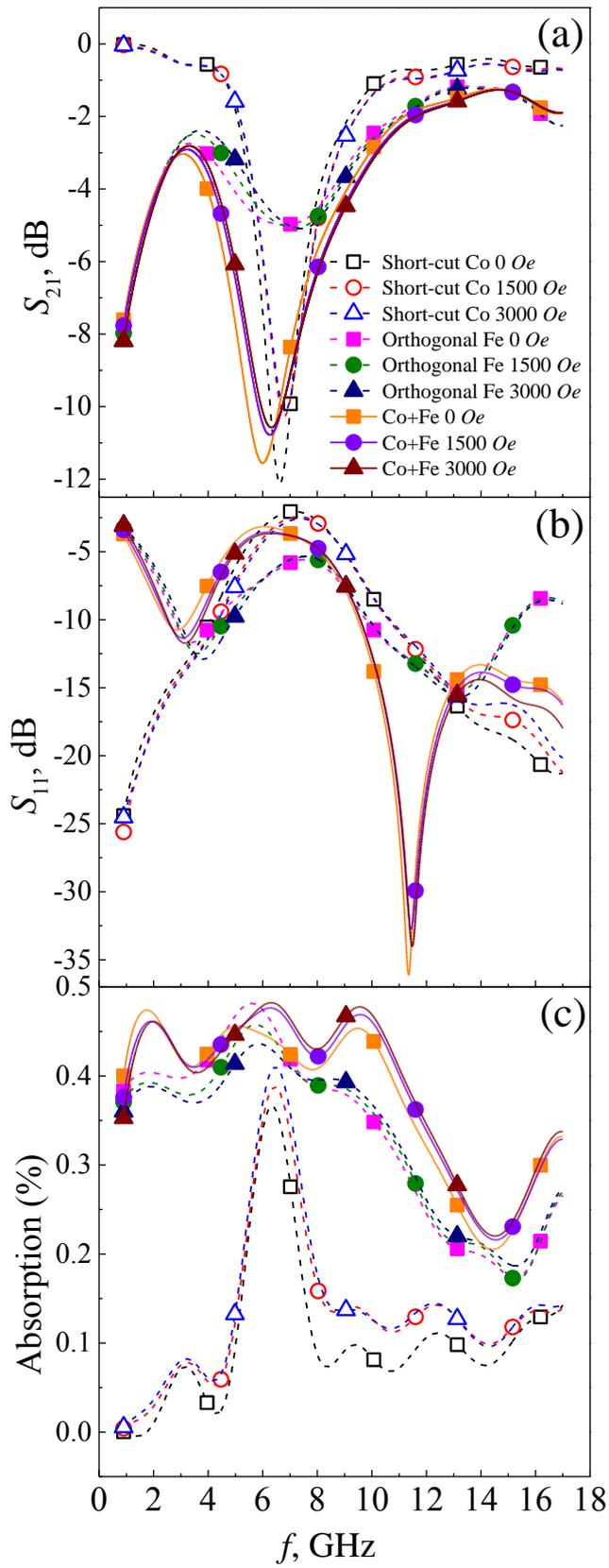

FIG 8

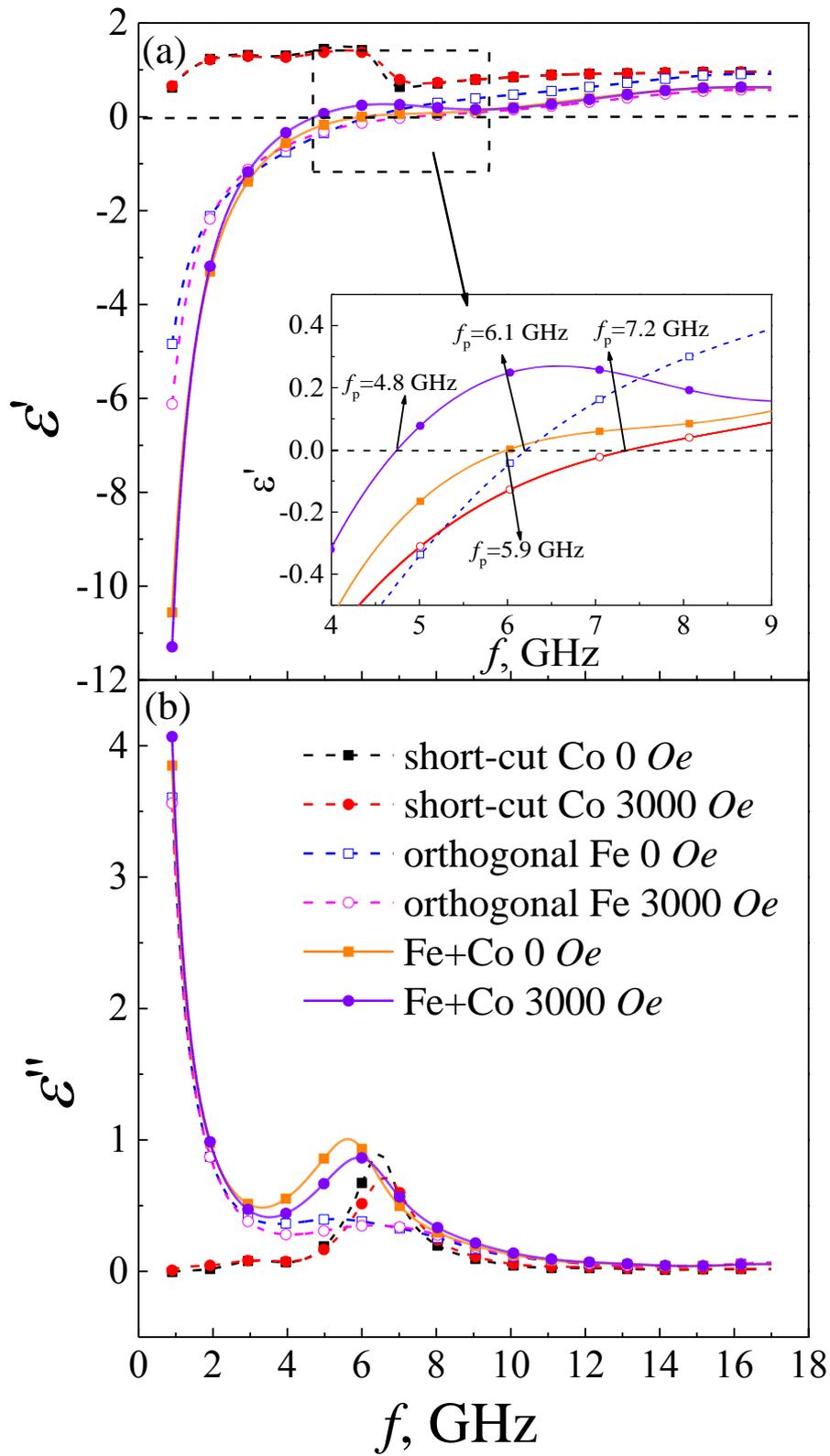

FIG 9